\documentclass[aps,twocolumn,nofootinbib,superscriptaddress]{revtex4}%
\usepackage{epsfig}
\usepackage{hyperref}
\usepackage{amsmath,amssymb,graphicx,epstopdf,enumerate,amsfonts,booktabs,color}
\usepackage{amsmath}
\usepackage{amsfonts}
\usepackage{amssymb}
\usepackage{graphicx}%
\providecommand{\U}[1]{\protect\rule{.1in}{.1in}}
\begin{document}
\title{ Universal Behaviors of Speed of Sound from Holography }
\author{Yi Yang}
\affiliation{Department of Electrophysics, National Chiao Tung University, Hsinchu, ROC}
\author{Pei-Hung Yuan}
\affiliation{Institute of Physics, National Chiao Tung University, Hsinchu, ROC}
\date{\today}

\begin{abstract}
We investigate the speed of sound in $(d+1)$-dimensional field theory by
studying its dual $\left(  d+2\right)  $-dimensional gravity theory from
gauge/gravity correspondence. Instead of the well known conformal limit
$c_{s}^{2}\rightarrow1/d$ at high temperature, we reveal two more universal
quantities in various limits: $c_{s}^{2}\rightarrow\left(  d-1\right)  /16\pi$
at low temperature and $c_{s}^{2}\rightarrow\left(  d-1\right)  /16\pi d$ at
large chemical potential.

\end{abstract}
\maketitle

\section{Introduction}

Gauge/gravity correspondence, as an useful tool to explore the field theories
with strong interaction, e.g. QCD, has been widely studied during the last
decade. However, one might hope to learn some qualitative lessons by searching
for quantities that do not depend on the details of the particular gravity
dual. Such 'universal' properties may apply to field theories without knowing
their gravity dual. An elegant example of such universal quantity is the ratio
$\eta/s$ of shear viscosity to entropy density. This takes the value $1/4\pi$
in all theories with gravity dual. In \cite{0905.0900,0905.0903}, it is showed
that the speed of sound approaches the conformal value $c_{s}^{2}=1/3$
universally from below in a general class of strongly interacting $\left(
3+1\right)  $-dimensional theories at high temperatures and zero chemical
potential. This result is consistent with the Monte Carlo lattice QCD
calculations \cite{0711.0656,1007.2580}. A number of string theory examples of
holographically dual theories, including both bottom-up models
\cite{0905.0900,0905.0903,0905.2969} and top-down models
\cite{0210220,0305064,0406200,0506002,0507026,0605076,0701132,0806.3796,0808.3953}
do indeed consistently show that $c_{s}^{2}\leq1/d$ case by case. The speed of
sound in QCD with isospin chemical potential \ has been studied in
\cite{0011365}, with the conjecture that the transition from hadron to quark
matter is smooth, it is showed that the speed of sound raises from $0$ to some
value close to $1$ (speed of light), then drops to some minimal value, and
then approaches $1/3$ from below at large isospin chemical potential. While
for baryon chemical potential, no physical systems in a deconfined phase with
a speed of sound exceeding the conformal value has prompted a conjecture that
this might represent a theoretical upper limit for the quantity.

In this work, we study $\left(  d+2\right)  $-dimensional
Einstein-Maxwell-Scalar (EMS) system, which has been widely studied as a
successful class of holographic QCD models by gauge/gravity correspondence. We
analytically obtained a general class of back-reacted black hole solutions and
study their dual $\left(  d+1\right)  $-dimensional field theories. We focused
on the behaviors of the speed of sound at arbitrary temperature and baryon
chemical potential. Instead of the well known conformal limit $c_{s}%
^{2}\rightarrow1/d$ at high temperature, we reveal two more universal
quantities in various limits: $c_{s}^{2}\rightarrow\left(  d-1\right)  /16\pi$
at low temperature and $c_{s}^{2}\rightarrow\left(  d-1\right)  /16\pi d$ at
large chemical potential.

We briefly review the EMS system and the solutions in section II. In section
III, we investigate the behavior of speed of sound in various limits. We
summarize our results in section IV.

\section{Einstein-Maxwell-Scalar Background}

To study holographic QCD theory in $\left(  d+1\right)  $-dimensional
spacetime, we consider a $\left(  d+2\right)  $-dimensional gravitational
background coupled to a Maxwell field and a neutral scalar field, i.e. the
Einstein-Maxwell-scalar system. In Einstein frame, the action is
\begin{align}
S  &  =\dfrac{1}{16\pi G_{d+2}}\int d^{d+2}x\sqrt{-g}\nonumber\\
&  \cdot\left[  R-{\frac{f\left(  {\phi}\right)  }{4}F^{2}}-\dfrac{1}%
{2}\left(  {\partial}_{\mu}{\phi}\right)  ^{2}-V\left(  \phi\right)  \right]
,
\end{align}
where $f\left(  {\phi}\right)  $ is a positive defined gauge kinematic
function. The equations of motion is derived as
\begin{align}
&  \nabla^{2}\phi=V_{\phi}+\frac{1}{4}f_{\phi}F^{2},\\
&  \nabla_{\mu}\left[  f F^{\mu\nu}\right]  ={{0,}}\\
&  R_{\mu\nu}-\frac{1}{2}g_{\mu\nu}R=\frac{f}{2}\left(  F_{\mu\rho}F_{\nu
}^{\rho}-\frac{1}{4}g_{\mu\nu}F^{2}\right) \nonumber\\
&  +\frac{1}{2}\left[  \partial_{\mu}\phi\partial_{\nu}\phi-\frac{1}{2}%
g_{\mu\nu}\left(  \partial\phi\right)  ^{2}-g_{\mu\nu}V\right]  .
\end{align}
To study the general asymptotic AdS black hole backgrounds with spherical
symmetry, we use the following ansatz for the fields,
\begin{align}
ds^{2}  &  =\frac{e^{2A\left(  z\right)  }}{z^{2}}\left[  -g(z)dt^{2}%
+\frac{dz^{2}}{g(z)}+d\vec{x}^{2}\right]  ,\label{metric}\\
\phi &  =\phi\left(  z\right)  \text{, \ }A_{\mu}=A_{t}\left(  z\right)  dt.
\end{align}
The equations of motion reduce to%
\begin{align}
\phi^{\prime\prime}+\left(  \frac{g^{\prime}}{g}+dw^{\prime}\right)
\phi^{\prime}+\frac{A_{t}^{\prime2}f_{\phi}}{2ge^{2w}}-\frac{e^{2w}V_{\phi}%
}{g}  &  =0,\\
A_{t}^{\prime\prime}+\left[  \frac{f^{\prime}}{f}+\left(  d-2\right)
w^{\prime}\right]  A_{t}^{\prime}  &  =0,\\
w^{\prime\prime}-w^{\prime2}+\dfrac{\phi^{\prime2}}{2d}  &  =0,\\
g^{\prime\prime}+dw^{\prime}g^{\prime}-\frac{fA_{t}^{\prime2}}{e^{2w}}  &
=0,\\
w^{\prime\prime}+dw^{\prime2}+\dfrac{3g^{\prime}}{2g}w^{\prime}+\dfrac
{g^{\prime\prime}+2e^{2w}V}{2dg}  &  =0,
\end{align}
where we defined $w\left(  z\right)  =A\left(  z\right)  -\ln z$.

Given the boundary conditions of regularity at the horizon $z=z_{H}$,%
\begin{equation}
g\left(  z_{H}\right)  =A_{t}\left(  z_{H}\right)  =0\text{,}%
\end{equation}
and asymptotic AdS spacetime at the boundary $z=0$,%
\begin{equation}
g(0)=f\left(  0\right)  =1,A\left(  0\right)  =A^{\prime}\left(  0\right)  =0,
\end{equation}
the most general black hole solutions can be analytically obtained as%
\begin{align}
\phi &  =\int_{0}^{z}\sqrt{2d\left(  w^{\prime2}-w^{\prime\prime}\right)
}dy,\label{phip}\\
A_{t}  &  =\mu\frac{\int_{z}^{z_{H}}\dfrac{e^{\left(  2-d\right)  w}}{f}%
dy}{\int_{0}^{z_{H}}\dfrac{e^{\left(  2-d\right)  w}}{f}dy}=\mu-\rho
z^{d-1}+\cdots,\label{At}\\
g  &  =1-\frac{\int_{0}^{z}e^{-dw}dy}{\int_{0}^{z_{H}}e^{-dw}dy}\nonumber\\
&  +\dfrac{\mu^{2}\left\vert
\begin{array}
[c]{cc}%
\int_{0}^{z_{H}}e^{-dw}dy & \int_{0}^{z_{H}}e^{-dw}dy\int_{0}^{y}%
\dfrac{e^{\left(  2-d\right)  w}}{f}dx\\
\int_{z_{H}}^{z}e^{-dw}dy & \int_{z_{H}}^{z}e^{-dw}dy\int_{0}^{y}%
\dfrac{e^{\left(  2-d\right)  w}}{f}dx
\end{array}
\right\vert }{\int_{0}^{z_{H}}e^{-dw}dz\left(  \int_{0}^{z_{H}}\dfrac
{e^{\left(  2-d\right)  w}}{f}dz\right)  ^{2}}\label{g}\\
V  &  =-\frac{e^{-2w}}{2}\left(  2dgw^{\prime\prime}+2d^{2}gw^{\prime
2}+3dg^{\prime}w^{\prime}+g^{\prime\prime}\right)  , \label{V}%
\end{align}
where $\mu$ is chemical potential and $\rho$ is the baryon density%
\begin{equation}
\rho=\dfrac{\mu}{\left(  d-1\right)  \int_{0}^{z_{H}}\dfrac{e^{\left(
2-d\right)  w}}{f}dy}. \label{rho}%
\end{equation}
In the solution Eq. (\ref{phip}-\ref{V}), the warped factor $w\left(
z\right)  $ and the gauge kinetic function $f\left(  \phi\right)  $ are two
arbitrary functions. We should note that, to guarantee the scalar field $\phi$
to be real, the warped factor $w\left(  z\right)  $ need to satisfy the
condition $w^{\prime2}\geq w^{\prime\prime}$, which leads to $A^{\prime\prime
}\left(  0\right)  \leq0$.

The entropy density and the temperature of the black hole can be obtained from
the background as
\begin{align}
s  &  =\dfrac{e^{dw\left(  z_{H}\right)  }}{4}.\label{s}\\
T  &  =\left\vert \dfrac{g^{\prime}\left(  z_{H}\right)  }{4\pi}\right\vert
=T_{0}\left[  1-\frac{\mu^{2}\int_{0}^{z_{H}}e^{-dw}dz\int_{z}^{z_{H}}%
\dfrac{e^{\left(  2-d\right)  w}}{f}dy}{\left(  \int_{0}^{z_{H}}%
\dfrac{e^{\left(  2-d\right)  w}}{f}dz\right)  ^{2}}\right]  , \label{T}%
\end{align}
where%
\begin{equation}
T_{0}=\frac{e^{-dw\left(  z_{H}\right)  }}{4\pi\int_{0}^{z_{H}}e^{-dw}dz},
\label{T0}%
\end{equation}
is the black hole temperature at $\mu=0$. It is easy to see that the high
temperature limit corresponds to $z_{H}\rightarrow0$.

\section{Speed of Sound}

In grand canonical ensemble with fixed chemical potential, the squared speed
of sound can be calculated as
\begin{equation}
c_{s}^{2}=\frac{s}{T\left(  \frac{\partial s}{\partial T}\right)  _{\mu}%
+\mu\left(  \frac{\partial\rho}{\partial T}\right)  _{\mu}},\label{cs}%
\end{equation}
Plugging Eqs. (\ref{s}-\ref{T}) into Eq. (\ref{cs}), it is straightforward to
obtain
\begin{equation}
c_{s}^{2}=\frac{c_{s0}^{2}+a\left(  1+c_{s0}^{2}\right)  \left[  1-b\left(
T_{0}-T\right)  \right]  \tilde{c}_{s0}^{2}}{1+a\left(  1+c_{s0}^{2}\right)
},\label{cs2}%
\end{equation}
where%
\begin{align}
\tilde{c}_{s0}^{2} &  =\frac{d-1}{16\pi}\text{,}\\
a &  =\frac{\left(  d-1\right)  \rho^{2}}{\pi T_{0}Tf\left(  z_{H}\right)
e^{2\left(  d-1\right)  w\left(  z_{H}\right)  }}\geq0\text{,}\label{a}\\
b &  =\frac{8\pi}{\left(  d-1\right)  \mu\rho\text{ }e^{-dw\left(
z_{H}\right)  }}\geq0,
\end{align}
and%
\begin{equation}
c_{s0}^{2}=-1-\frac{e^{-dw\left(  z_{H}\right)  }}{dw^{\prime}\left(
z_{H}\right)  \left(  \int_{0}^{z_{H}}e^{-dw}dz\right)  },\label{cs0}%
\end{equation}
is the\ squared speed of sound at $\mu=0$.

\noindent\textbf{Zero Chemical Potential.} First, we will study the properties
for the speed of sound at zero chemical potential, i.e. $\mu=0$. It is well
known that the squared speed of sound\ in QCD approaches the conformal limit
$1/3$ in high temperature limit. From Eq. (\ref{cs0}), it is easy to show
that,
\begin{equation}
\lim_{T\rightarrow\infty}c_{s0}^{2}=\lim_{z_{H}\rightarrow0}\left[
-1-\frac{e^{-dw\left(  z_{H}\right)  }}{dw^{\prime}\left(  z_{H}\right)
\left(  \int_{0}^{z_{H}}e^{-dw}dz\right)  }\right]  =\frac{1}{d},
\label{conformal}%
\end{equation}
is an universal quantity that is model independent, i.e. independent of the
choice of the functions $w\left(  z\right)  $ and $f\left(  z\right)  $ in the
above solution Eqs.(\ref{phip}-\ref{V}). Eq. (\ref{conformal}) applies to
$\left(  d+1\right)  $-dimensional QCD, which is the natural generalization of
the $\left(  3+1\right)  $-dimensional QCD.

To further investigate the behavior of the speed of sound in high temperature
limit, we expand $c_{s0}^{2}$ at $z_{H}=0$,
\begin{equation}
c_{s0}^{2}=\frac{1}{d}+\frac{3\left(  d+1\right)  }{d\left(  d+3\right)
}A_{0}^{\prime\prime}z_{H}^{2}+O\left(  z_{H}^{4}\right)  . \label{cs0T}%
\end{equation}
Since $A^{\prime\prime}\left(  0\right)  \leq0$, the coefficient of $z_{H}%
^{2}$ in Eq. (\ref{cs0T}) is always negative. It indicates that, in high
temperature limit, the squared speed of sound approaches to $1/d$ from below.

To obtain the bound for the speed of sound at arbitrary temperature, we
calculate the extreme of speed of sound by differentiating it with respect to
$z_{H}$,
\begin{equation}
\frac{dc_{s0}^{2}}{dzH}=\frac{e^{-dw\left(  z_{H}\right)  }}{\int_{0}^{z_{H}%
}e^{-dw}dz}\left(  \frac{w^{\prime\prime}\left(  z_{H}\right)  }{dw^{\prime
2}\left(  z_{H}\right)  }-c_{s0}^{2}\right)  =0,
\end{equation}
which leads to%
\begin{equation}
c_{s0}^{2}=\frac{w^{\prime\prime}\left(  z_{H}\right)  }{dw^{\prime2}\left(
z_{H}\right)  }\leq\frac{1}{d},\label{bound}%
\end{equation}
in which we have used the condition $w^{\prime2}\geq w^{\prime\prime}$.

We therefore conclude that, at zero chemical potential $\mu=0$, the squared
speed of sound approaches to its maximum value, the conformal limit $1/d$, in
high temperature limit for $\left(  d+1\right)  $-dimensional QCD. This is
consistent with the recent results from lattice QCD \cite{Phy.Rep.}.

\noindent\textbf{Finite Chemical Potential. }Next, we will study the
properties for the speed of sound at finite chemical potential. i.e.
$0<\mu<\infty$. In high temperature limit, $T,T_{0}\rightarrow\infty$,
$a\rightarrow0$. By using Eq. (\ref{cs2}), it is easy to verify that
\begin{equation}
\lim_{T\rightarrow\infty}c_{s}^{2}=\lim_{T\rightarrow\infty}c_{s0}^{2}%
=\frac{1}{d}. \label{highT}%
\end{equation}
Thus the speed of sound\ in $\left(  d+1\right)  $-dimensional QCD approaches
to the same universal value $1/d$ in high temperature limit even for finite
chemical potential.

Similarly, we expand $c_{s}^{2}$ at $z_{H}=0$,%
\begin{align}
c_{s}^{2}  &  =\frac{1}{d}+\left[  \frac{3\left(  d+1\right)  }{d\left(
d+3\right)  }A_{0}^{\prime\prime}+\frac{\mu^{2}\left(  d-1\right)  ^{2}}%
{d^{2}\left(  d+1\right)  }\left(  1-\frac{1}{\tilde{c}_{s0}^{2}}\right)
\right]  z_{H}^{2}\nonumber\\
&  +O\left(  z_{H}^{4}\right)  . \label{csT}%
\end{align}
Since $A^{\prime\prime}\left(  0\right)  \leq0$, the coefficient of $z_{H}%
^{2}$ in Eq. (\ref{csT}) is always negative\ for $d<1+16\pi\simeq52$. This
indicates that, in high temperature limit, the speed of sound also approaches
to $1/d$ from below for finite chemical potential.

Furthermore, it is easy to see that $T<T_{0}$ from Eq. (\ref{T}), which
implies, from Eq. (\ref{cs2}), that
\begin{equation}
c_{s}^{2}\leq\frac{c_{s0}^{2}+a\left(  1+c_{s0}^{2}\right)  \tilde{c}_{s0}%
^{2}}{1+a\left(  1+c_{s0}^{2}\right)  }\leq\max\left(  c_{s0}^{2},\tilde
{c}_{s0}^{2}\right)  .
\end{equation}
In Eq. (\ref{bound}), we have proved $c_{s0}^{2}<1/d$, and it is easy to show
that $\tilde{c}_{s0}^{2}=\left(  d-1\right)  /16\pi<1/d$ for $d\leq7$. We thus
conclude that, at finite chemical potential, the squared speed of sound
approaches to its maximum value, the conformal limit $1/d$, in high
temperature limit for $\left(  d+1\right)  $-dimensional QCD at least for
$d\leq7$.

\noindent\textbf{Large Chemical Potential. }For large chemical potential, a
new phase, color-flavor-locking or color superconductivity, has been
conjectured in QCD. It is thus interesting to investigate the behavior of
speed of sound at large chemical potential. By using Eq. (\ref{rho}) and Eq.
(\ref{T}), we rewrite the speed of sound Eq. (\ref{cs2}) in the following
form,
\begin{equation}
c_{s}^{2}=\frac{c_{s0}^{2}+a\left(  1+c_{s0}^{2}\right)  \left[  1-\frac
{2\int_{0}^{z_{H}}e^{-dw}dz\int_{z}^{z_{H}}\dfrac{e^{\left(  2-d\right)  w}%
}{f}dy}{\int_{0}^{z_{H}}e^{-dw}dz\int_{0}^{z_{H}}\dfrac{e^{\left(  2-d\right)
w}}{f}dz}\right]  \tilde{c}_{s0}^{2}}{1+a\left(  1+c_{s0}^{2}\right)  }.
\end{equation}
We should be more careful to take the large chemical potential limit because
that, from Eq. (\ref{T}), the temperature become negative when $\mu$ exceed a
critical value, which certainly does not make sense physically. To take the
large chemical potential limit at a fixed temperature, the correct process is
to take the double limits of $\mu\rightarrow\infty$ with $z_{H}\rightarrow0$
together. Under the double limits, $a\rightarrow\infty$\ and the speed of
sound reduces to
\begin{equation}
\lim_{\substack{\mu\rightarrow\infty\\z_{H}\rightarrow0}}c_{s}^{2}%
=\frac{\tilde{c}_{s0}^{2}}{d}=\frac{d-1}{16\pi d}.\label{csmu}%
\end{equation}
In stead of the conformal limit $c_{s}^{2}\rightarrow1/d$ in high temperature
limit, we find another universal quantity $c_{s}^{2}\rightarrow\left(
d-1\right)  /16\pi d$ in the limit of infinity chemical potential. Eq.
(\ref{csmu}) shows that the speed of sound does not approach to the conformal
limit at large chemical potentia., This implies that the theory at large
chemical potential is in a new phase which is different from the phase at high
temperature as people conjectured.

\noindent\textbf{Low Temperature. }We further analyze the behavior of speed of
sound in low temperature limit $T\rightarrow0$ which leads to $a\rightarrow
\infty$. From Eq. (\ref{cs2}), we have%
\begin{equation}
\lim_{T\rightarrow0}c_{s}^{2}=\tilde{c}_{s0}^{2}=\frac{d-1}{16\pi
}.\label{lowT}%
\end{equation}
Remarkably, we find one more universal quantity $c_{s}^{2}\rightarrow\left(
d-1\right)  /16\pi$ in low temperature limit.

\begin{figure}[t]
\begin{center}
\includegraphics[
height=2in, width=2.6in]
{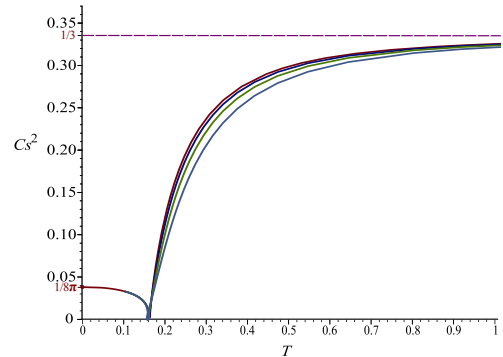}
\end{center}
\caption{squared speed of sound v.s. temperature at $\mu=0, 0.4779, 0.08,
0.12$ from upper to lower lines.}%
\label{figT}%
\end{figure}

\begin{figure}[t]
\begin{center}
\includegraphics[
height=2in, width=2.6in]
{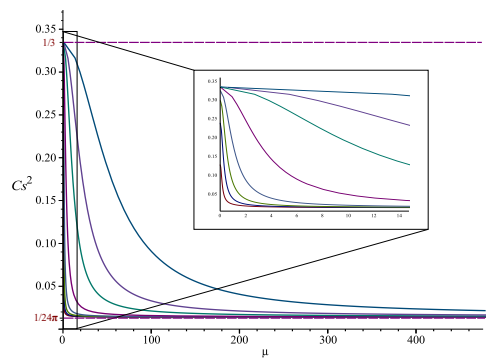}
\end{center}
\caption{squared speed of sound v.s. chemical potential at $T=0.2, 0.3, 0.5,
1, 3, 10, 20, 50$ from lower to upper lines.}%
\label{figmu}%
\end{figure}

Finally, we present an explicit example of $(3+1)$-dimensional holographic QCD
model by taking the warped factor $w\left(  z\right)  $ and gauge kinetic
function $f\left(  \phi\right)  $ in \cite{1703.09184}. For $d=3$, the
universal quantities that we found in Eqs. (\ref{highT},\ref{csmu},\ref{lowT})
reduce to
\begin{align}
\text{at large }T &  :c_{s}^{2}=\frac{1}{d}\rightarrow\frac{1}{3},\\
\text{at small }T &  :c_{s}^{2}=\frac{d-1}{16\pi}\rightarrow\frac{1}{8\pi},\\
\text{at large }\mu &  :c_{s}^{2}=\frac{d-1}{16\pi d}\rightarrow\frac{1}%
{24\pi}.
\end{align}
We plot the squared speed of sound v.s. temperature and chemical potential in
Fig.\ref{figT} and Fig.\ref{figmu} respectively. The behaviors of the sound
speed in the figures perfectly match the analysis in this work.

\section{Summary}

In this work, we studied gauge/gravity correspondence by considering the
$\left(  d+2\right)  $-dimensional Einstein-Maxwell-Scalar system, which has
been widely studied as a successful class of holographic QCD models. We
analytically obtained a general class of back-reacted solutions and focused on
the behaviors of speed of sound at arbitrary temperature and chemical
potential in $\left(  d+1\right)  $-dimensional holographic QCD. We found
that, in various limits, the speed of sound approaches certain universal
quantities which are not dependent on the details of the models. The universal
behaviors of speed of sound we have found in this work are follows:

\begin{itemize}
\item $c_{s}^{2}\rightarrow\frac{1}{d}$, as $T\rightarrow\infty$ with fixed
$\mu$; 

\item $c_{s}^{2}\rightarrow\frac{d-1}{16\pi}$, as $T\rightarrow0$ with fixed
$\mu$;

\item $c_{s}^{2}\rightarrow\frac{d-1}{16\pi d}$, as $\mu\rightarrow\infty$
with fixed $T$.
\end{itemize}

We also proved that $c_{s}^{2}\leq1/d$ for all temperature and chemical
potentials (at least for $d\leq7$) provided\ that the solution of scalar
$\phi$ in Eq. (\ref{phip}) is real.

To investigate these universal quantities in further details is not only
attractive in QCD theory, such as the new phase in large chemical potential,
but also important to understand the deep structure of the gauge/gravity correspondence.

\begin{acknowledgments}
We would like to thank Carlos Hoyos for useful discussions. This work is
supported in part by the Ministry of Science and Technology and S.T. Yau
center at NCTU, Taiwan.
\end{acknowledgments}

\end{document}